# Observational Confirmation of the Sun's CNO Cycle

Michael Mozina[1], Hilton Ratcliffe[2] and O. Manuel[3]

___________________________________

Measurements on γ-rays from a solar flare in Active Region 10039 on 23 July 2002 with the RHESSI spacecraft spectrometer indicate that the CNO cycle occurs at the solar surface, in electrical discharges along closed magnetic loops. At the two feet of the loop, $H^+$ ions are accelerated to energy levels that surpass Coulomb barriers for the $^{12}C(^{1}H, \gamma)^{13}N$ and $^{14}N(^{1}H, \gamma)^{15}O$ reactions. First x-rays appear along the discharge path. Next annihilation of $\beta^+$-particles from $^{13}N$ and $^{15}O$ (t½ = 10 m and 2 m) produce bright spots of 0.511 MeV γ's at the loop feet. As $^{13}C$ increases from $\beta^+$-decay of $^{13}N$, the $^{13}C(\alpha, n)^{16}O$ reaction produces neutrons and then the 2.2 MeV emission line appears from n-capture on $^{1}H$. These results suggest that the CNO cycle changed the $^{15}N/^{14}N$ ratio in the solar wind and at the solar surface over geologic time, and this ratio may contain an important historical record of climate changes related to sunspot activity.

___________________________________

[1] President, Emerging Technologies, P. O. Box 1539, Mt. Shasta, CA 96067, USA, *michael@etwebsite.com*
1-800-729-4445
[2] Astronomical Society of South Africa, PO Box 354, Kloof 3640, South Africa, ratcliff@iafrica.com
[3] Nuclear Chemistry, University of Missouri, Rolla, MO 65401 USA, omatumr@yahoo.com

## I.     INTRODUCTION

Deep-seated magnetic fields accelerate H$^+$ ions, an ionized neutron-decay product, upward from the Sun's core [1]. These protons act as the "carrier gas" that maintains mass separation in the Sun, covering its surface with lightweight elements [1]. Until recently it was widely assumed that H-fusion generates stellar luminosity and that H, He, C, N, and other light elements are plentiful inside ordinary stars. Since the probability of four hydrogen atoms fusing into a helium atom is small, the late Hans Bethe [2] proposed in 1939 that $^{12}$C serves as a catalyst for the fusion of hydrogen into helium *via* the CNO cycle in the core of the Sun:

1. $^{12}$C + $^{1}$H → $^{13}$N + γ
2. $^{13}$N → $^{13}$C + β$^+$ + γ + ν
3. $^{13}$C + $^{1}$H → $^{14}$N + γ
4. $^{14}$N + $^{1}$H → $^{15}$O + γ
5. $^{15}$O → $^{15}$N + β$^+$ + γ + ν
6. $^{15}$N + $^{1}$H → $^{12}$C + $^{4}$He

At solar temperatures, each of the above atoms is likely a positive ion. Positrons (β$^+$) emitted by the decay of $^{13}$N and $^{15}$O in steps 2 and 5 will react with electrons to release the 0.511 MeV γ–rays characteristic of annihilation. Electrical fields that accelerate H$^+$ ions to energies that permit the occurrence of reactions 1, 3, 4 and 6 may also accelerate He$^{++}$ ions to energies that destroy $^{13}$C [3] in a process that competes with reaction 3 and then goes on to generate $^{2}$H instead of the $^{15}$O product shown in reaction 4. These competing reactions are shown below as 3' and 4':

3'. $^{13}$C + $^{4}$He → $^{16}$O + $^{1}$n
4'. $^{1}$n + $^{1}$H → $^{2}$H + γ

Burbidge *et al.* [3] suggested reaction 3' as a process that generates neutrons inside stars. Neutrons released into H-rich material would likely be captured by $^1$H and might be detected by observing the 2.223 MeV γ released in reaction 4'.

Neutrinos (ν) emitted in the decay of $^{13}$N and $^{15}$O in steps 2 and 5 may exceed the 0.86 MeV threshold of the $^{37}$Cl solar neutrino detector that Ray Davis proposed in 1955 [4]. The embarrassingly low flux of solar neutrinos found in all solar neutrino measurements [e.g., 5, 6] convinced the scientific community that a.) the proton-proton chain, with $E_\nu \leq 0.41$ MeV, is the main source of solar energy, and b.) Bethe's CNO cycle produces little, if any, of the Sun's energy.

However other quantitative measurements on the Sun revealed puzzling hints that a solar CNO cycle operates near the solar surface, where H, He, C and N are abundant [1], rather than in the Sun's interior. Rare isotopes of carbon and nitrogen, $^{13}$C and $^{15}$N, are produced by reactions 2 and 5 in the CNO cycle outlined above. In 1975 Kerridge [7] noted that the $^{15}$N/$^{14}$N ratio in the solar wind appears to have increased over geologic time. The ancient solar wind and modern solar flares release nitrogen with less $^{15}$N than is in the modern solar-wind nitrogen [8].

The $^{15}$N/$^{14}$N ratio in the solar wind has not steadily increased with time. Like sunspot activity at the solar surface, the $^{15}$N/$^{14}$N ratio in the solar wind exhibits evidence of large, sporadic changes [9]. A secular increase in the $^{13}$C/$^{12}$C ratio in the solar wind correlates with the increase in the $^{15}$N/$^{14}$N ratio [10, 11], as expected by the addition of products from reactions 2 and 5 of the above CNO cycle.

Here are a few other pertinent but unexpected experimental findings on the Sun:
   a. Lightweight isotopes (**L**) of many elements are enriched relative to heavier ones (**H**) in the solar wind, as if each element had passed through 9 theoretical stages of mass fractionation, each enriching the (**L/H**) ratio by (**H/L**)$^{0.5}$ [12].

- 3 -

b. The lightweight isotopes (**L**) of most elements are systematically less enriched relative to heavier ones (**H**) in solar flares, as if these violent surface events by-passed about 3.5 of the 9 theoretical stages of mass fractionation [13].

c. The behavior of nitrogen isotopes in the solar wind and in solar flares is opposite to those of other elements. For nitrogen **L/H** = $^{14}N/^{15}N$, and the value of this ratio is higher in solar flares than in the quiet solar wind [8,13].

d. In 1977 solar-induced variations in the geomagnetic field first hinted that the Sun might be a pulsar [14] that formed on the collapsed core of a supernova [15]. Earlier this year Mozina [16, 17] discovered rigid, iron-rich structures below the Sun's fluid photosphere, and helio-seismology data have since confirmed that the Sun is stratified at relatively shallow depths beneath the visible photosphere, at ≈ 0.5% solar radii (≈0.005 $R_o$) [18].

e. As mentioned earlier, the Sun is a magnetic plasma diffuser that maintains mass separation by an upward flow of the ionized neutron-decay product ($H^+$ ions) coming from the solar core [1]. Fusion consumes most $H^+$ ions in their upward journey along deep-seated magnetic fields from the core of the Sun and generates <38% of the Sun's energy [19]. At the solar surface these magnetic fields may continue upward or form closed loops in active regions where solar flares and eruptions occur. The $H^+$ ions are accelerated to high energies in the magnetic loops shown in Fig. 1, generating an *"electrified gas"* that heats the corona [20].

The next section will show how the results of γ-ray spectrometry on the RHESSI spacecraft is the key that unlocks the mystery of several of these puzzling solar observations and reveals new details of the Sun's operation and the location of its CNO cycle [2, 3].

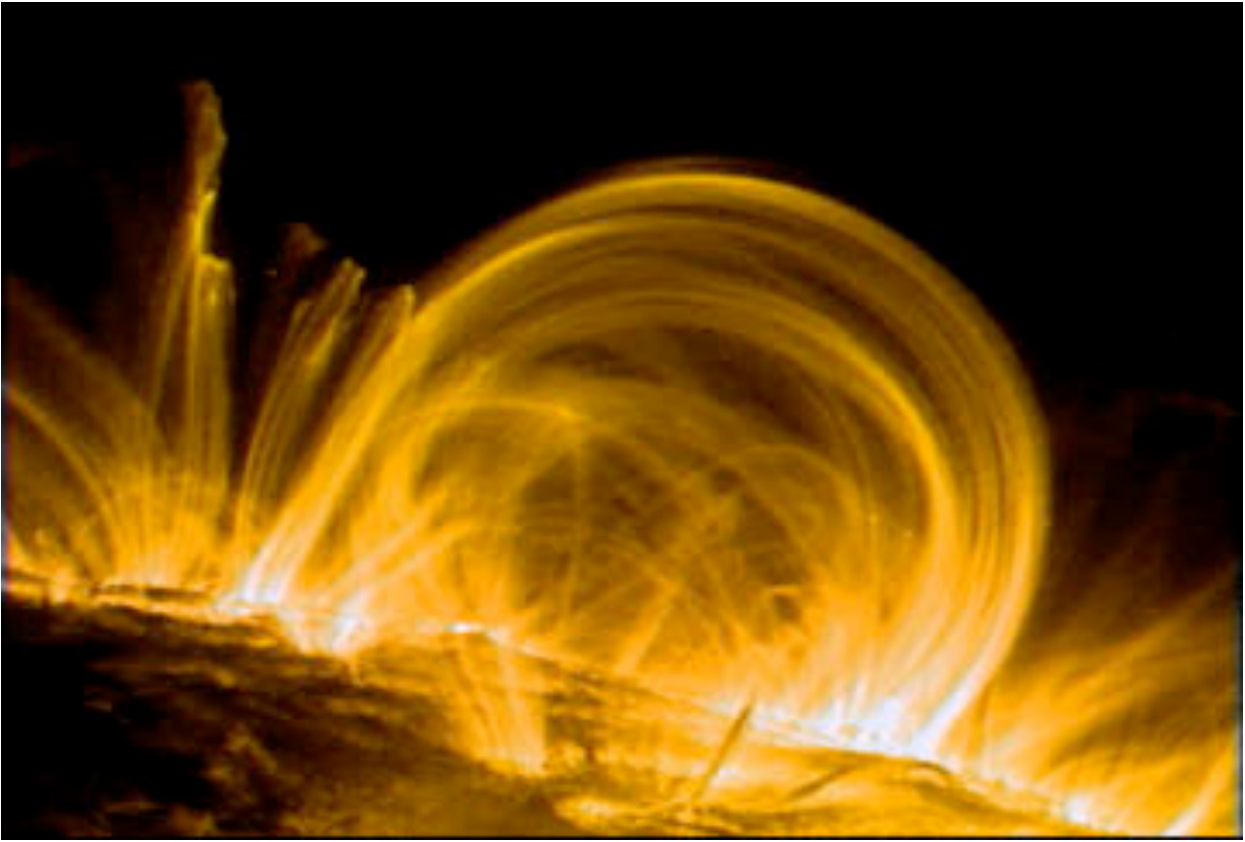

**Fig. 1.** This is a false color image taken with NASA's TRACE spacecraft of ultraviolet light emitted as loops of electrified gas are *"heated to temperatures 300 times greater than the Sun's visible surface"* [20]. The most intense heating (white regions) occurs at the base of the magnetic loops, where the fields emerge from and return to the solar surface. γ-Ray spectroscopy of another flare event with the RHESSI spacecraft [21] reveals annihilation of the positrons made in the discharge loops by steps 2 and 5 of the CNO cycle and capture of the neutrons made by the competing reactions, $^{13}C(\alpha, n)^{16}O$ and $^{1}H(n, \gamma)^{2}H$, shown above as steps 3' and 4'.

## II. NEW EXPERIMENTAL OBSERVATIONS

The RHESSI spacecraft was launched on 5 Feb 2002 for the purpose of studying the process of particle acceleration and energy release in solar flares. The spectrometer on board is designed to provide simultaneous, high-resolution imaging and spectroscopy of solar flares, from 3 keV X-rays to 17 MeV γ-rays with high time resolution. Fig 2 shows four sequential time frames from NASA's animation of spectrometry measurements on the solar flare event at Active Region 10039 in the early morning of 23 July 2002 [21]. These cover a time span of 10 min and 9 sec.

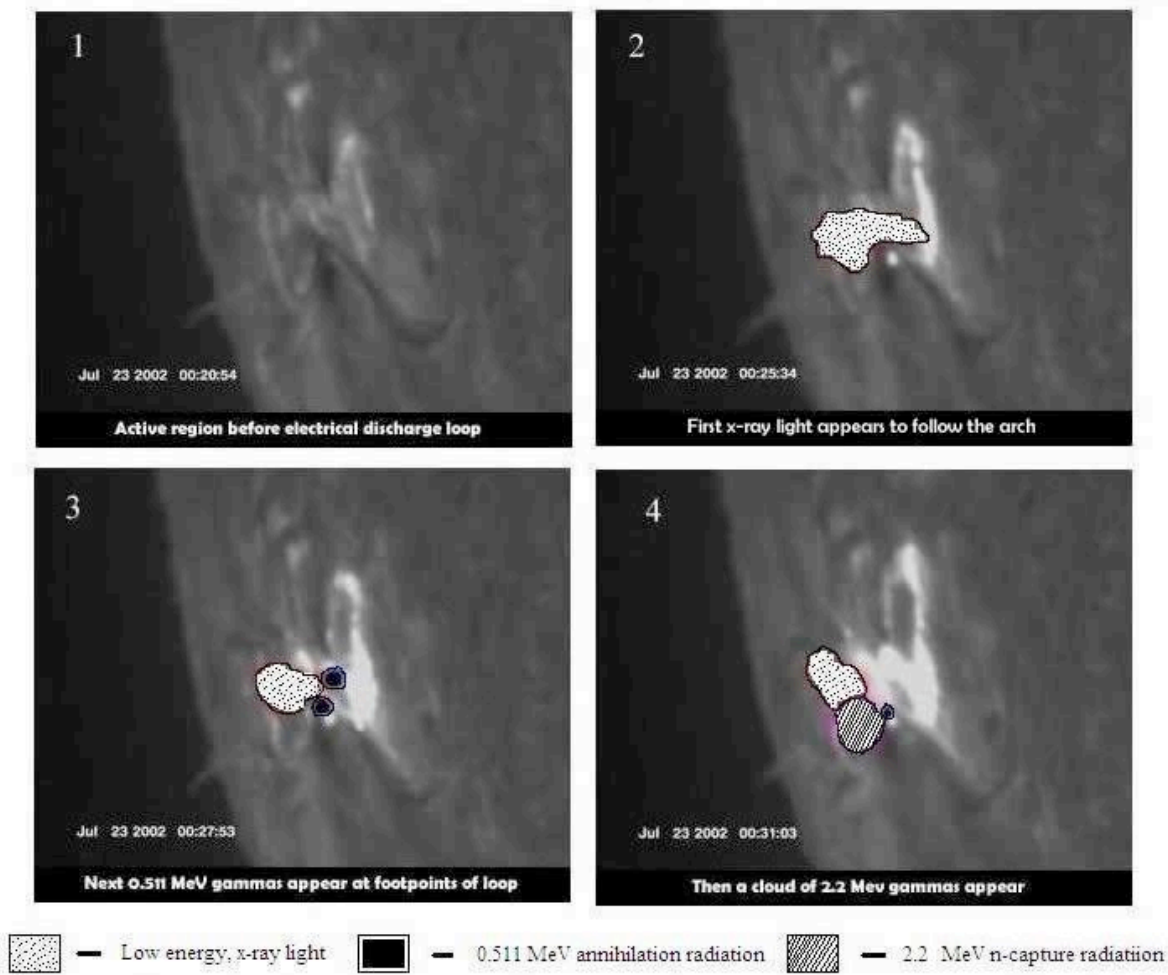

**Fig. 2.** This shows four sequential animation frames of the solar flare event at Active Region 10039 recorded with a spectrometer on the RHESSI spacecraft on 23 July 2002 [21]. Frame 1 shows the area at 00:20:54. Frame 2 shows the area 4 min and 40 sec later, at 00:25:34, when 12-25 keV x-rays appear along the discharge loop. Frame 3 shows the area almost 7 min after the first frame, at 00:27:53, when 0.511 MeV γ's appear as dark footpoints of the discharge loop. Frame 4 shows the area 10 min and 9 sec after the first frame, when 2.223 MeV γ's reveal a region where hydrogen is undergoing neutron-capture.

The appearance and disappearance of different light sources in the flare event may be better seen in the original animation, http://svs.gsfc.nasa.gov/vis/a000000/a002700/a002750/. There low-energy emissions (12-25 keV x-rays) emerge first in red along the magnetic discharge loop, next the 0.511 MeV annihilation γ's appear as two blue footpoints of the discharge loop, and then the 2.223 MeV neutron-capture γ's appear later as a violet cloud above the footpoints.

The sequence and location of these light emissions are consistent with those expected from the occurrence of the CNO reactions shown above. First, the x-rays likely appear when highly ionized chemical species form along the discharge loop. $H^+$ ions may be accelerated in the loop to energy levels that surpass Coulomb barriers for the $^{12}C(^1H, γ)^{13}N$ and $^{14}N(^1H, γ)^{15}O$ reactions at the feet of the loop. These products have $β^+$-decay half-lives of 10 m and 2 m, respectively. There is thus a delay in the appearance of the 0.511 MeV γ's from $β^+$-annihilation reactions at the loop feet. There is an additional delay in the emission of 2.223 MeV γ's from neutron-capture reactions. For this reaction to occur, $^{13}N$ nuclei (t½ = 10 m) must first decay to $^{13}C$. The $^{13}C$ nuclei are stable and may increase in concentration and then interact with $^4He^{++}$ ions (α particles) to produced neutrons via the $^{13}C(α, n)^{16}O$ reaction. The neutrons have an 11 min half-life and will reasonably build up to some maximum concentration where the rates of production and decay are balanced. This would likely correspond to maximum intensity of the 2.223 MeV γ's from neutron-capture on hydrogen.

III. CONCLUSIONS

The above findings [7-21] suggest that Bethe's solar CNO cycle [2] has made $^{13}N$, $^{13}C$, $^{15}O$ and $^{15}N$ at the surface of the Sun over geologic time [7-11] and now makes these unstable or rare isotopes in electrical discharge loops of solar flares [21]. Temporal changes in sunspot activity likely explain variations in the solar $^{15}N/^{14}N$ ratio. If light elements like H, C, N and O had not moved selectively to the solar surface [12, 13, 17, 19], H-fusion *via* the CNO cycle [2] might have occurred deep in the Sun. We look forward to other explanations for these findings [7-21].

**ACKNOWLEDGEMENTS**

Support from the University of Missouri-Rolla and the Foundation for Chemical Research, Inc. (FCR) are gratefully acknowledged. We are grateful to the scientists - Drs. Robert Lin, Sam Krucker, Gordon J. Hurford, and David M. Smith (University of California at Berkeley), Drs. R. J. Murphy and G. H. Share (NRL), Dr. X.-M. Hua (L-3 Communications Analytics Corporation), Dr. Richard A. Schwartz (NASA/ GSFC), and Dr. Benzion Kozlovsky (Tel Aviv University) - for allowing spectroscopic data of the 23 July 2002 solar flare event to be animated and posted at http://svs.gsfc.nasa.gov/vis/a000000/a002700/a002750/ . The results are shown in an abbreviated form in Fig. 2
**REFERENCES**

1. O. K. Manuel, B. W. Ninham and S. E. Friberg, *J. Fusion Energy*, **21**, 193-198 (2002).

2. Hans Bethe, *Phys. Rev.,* **55**, 103 (1939).

3. E. M. Burbidge, G. R. Burbidge, W. A. Fowler and F. Hoyle, *Rev. Mod. Phys.,* **29**, 547-650 (1957).

4. R. Davis, Jr., *Phys. Rev.,* **97**, 766-769 (1955).

5. R. Davis, Jr., D. S. Harmer and K. C. Hoffman, *Phys. Rev. Lett.,* **20**, 1205-1209 (1968).

6. Q. R. Ahmad, et al., *Phys. Rev. Lett.,* **89**, 011301, 6 pp. (2002).

7. J. F. Kerridge, *Science,* **188**, 162-164 (1975).

8. J. F. Kerridge, *Rev. Geophys.* **31**, 423-437 (1993).

9. J. S. Kim, Y. Kim, K. Marti and J. F. Kerridge, *Nature*, **375**, 383-385 (1995).

10. R. H. Becker, *Earth Planet. Sci. Lett.,* **50**, 189-196 (1980).

11. J. Geiss and P. Boschler, *Geochim. Cosmochim. Acta*, **46**, 529-548 (1982).

12. O. Manuel and G. Hwaung, *Meteoritics,* **18**, 209-222 (1983).

13. O. Manuel, *in* Oliver K. Manuel (Ed), *Proceedings of the 1999 ACS Symposium on the Origin of Elements in the Solar System: Implications for Post-1957 Observations* (Klurwer/Plenum Publishers, NY, pp. 279-287, 2000).

14. P. Toth, *Nature,* **270**, 159-160 (1977).

15. O. K. Manuel and D. D. Sabu, *Science,* **195**, 208-209 (1977).

16. M. Mozina, "The surface of the Sun", http://www.thesurfaceofthesun.com/index.html

17. O. Manuel, S. Kamat, and M. Mozina, *in* Eric J. Lerner and Jose B. Almeida (Eds), *Proceedings First Crisis in Cosmology Conf.* (AIP, Melville, NY, in press, 2005) http://arxiv.org/abs/astro-ph/0510001

18. S. Lefebvre and A. Kosovichev, "Changes in subsurface stratification of the Sun with the 11-year activity cycle", *Ap. J.,* **633**, L149-L (2005). http://xxx.lanl.gov/pdf/astro-ph/0510111

19. O. Manuel, E. Miller and A Katragada, *J. Fusion Energy*, **20**, 197-201 (2001).

20. NASA, "Fountains of fire illuminate solar mystery, overturn 30 year old theory", http://www.gsfc.nasa.gov/gsfc/spacesci/sunearth/tracecl.htm

21. W. Steigerwald, "RHESSI observes 2.2 MeV line emission from a solar flare", *in* SVS Animation 2750, http://svs.gsfc.nasa.gov/vis/a000000/a002700/a002750/